%Paper: hep-ph/9411297
%From: Ken-ichi Hikasa <hikasa@tuhep.phys.tohoku.ac.jp>
%Date: Tue, 15 Nov 94 15:36:13 JST
%Date (revised): Fri, 18 Nov 94 18:17:57 JST

% use plain.tex
%%%%%%%%%%%%%%%%%%%%%%%%%%%%%%%%%%%%%%%%%%%%%%%%%%%%%%%%%%%%%%%%%%%%%%%%%%
\catcode`\@=11

\font\fourteenrm=cmr10 scaled\magstep2
\font\twelverm=cmr10 scaled\magstep1
\font\ninerm=cmr9	     \font\sixrm=cmr6
\font\seventeenbf=cmbx10 scaled\magstep3
\font\fourteenbf=cmbx10 scaled\magstep2
\font\twelvebf=cmbx10 scaled\magstep1
\font\ninebf=cmbx9	      \font\sixbf=cmbx6
\font\seventeeni=cmmi10 scaled\magstep3	    \skewchar\seventeeni='177
\font\fourteeni=cmmi10 scaled\magstep2	    \skewchar\fourteeni='177
\font\twelvei=cmmi10 scaled\magstep1	    \skewchar\twelvei='177
\font\ninei=cmmi9			    \skewchar\ninei='177
\font\sixi=cmmi6			    \skewchar\sixi='177
\font\seventeensy=cmsy10 scaled\magstep3    \skewchar\seventeensy='60
\font\fourteensy=cmsy10 scaled\magstep2	    \skewchar\fourteensy='60
\font\twelvesy=cmsy10 scaled\magstep1	    \skewchar\twelvesy='60
\font\ninesy=cmsy9			    \skewchar\ninesy='60
\font\sixsy=cmsy6			    \skewchar\sixsy='60

\font\fourteenex=cmex10 scaled\magstep2
\font\twelveex=cmex10 scaled\magstep1

\font\fourteensl=cmsl10 scaled\magstep2
\font\twelvesl=cmsl10 scaled\magstep1
\font\ninesl=cmsl9

\font\fourteenit=cmti10 scaled\magstep2
\font\twelveit=cmti10 scaled\magstep1
\font\twelvett=cmtt10 scaled\magstep1
\font\twelvecp=cmcsc10 scaled\magstep1
\font\tencp=cmcsc10
\newfam\cpfam
\newcount\f@ntkey	     \f@ntkey=0
\def\samef@nt{\relax \ifcase\f@ntkey \rm \or\oldstyle \or\or
	 \or\it \or\sl \or\bf \or\tt \or\caps \fi }
\def\fourteenpoint{\relax
   \textfont0=\fourteenrm	    \scriptfont0=\tenrm
   \scriptscriptfont0=\sevenrm
   \def\rm{\fam0 \fourteenrm \f@ntkey=0 }%
   \textfont1=\fourteeni	    \scriptfont1=\teni
   \scriptscriptfont1=\seveni
   \def\oldstyle{\fam1 \fourteeni\f@ntkey=1 }%
   \textfont2=\fourteensy	    \scriptfont2=\tensy
   \scriptscriptfont2=\sevensy
   \textfont3=\fourteenex     \scriptfont3=\fourteenex
   \scriptscriptfont3=\fourteenex
   \def\it{\fam\itfam \fourteenit\f@ntkey=4 }\textfont\itfam=\fourteenit
   \def\sl{\fam\slfam \fourteensl\f@ntkey=5 }\textfont\slfam=\fourteensl
   \scriptfont\slfam=\tensl
   \def\bf{\fam\bffam \fourteenbf\f@ntkey=6 }\textfont\bffam=\fourteenbf
   \scriptfont\bffam=\tenbf	 \scriptscriptfont\bffam=\sevenbf
   \def\tt{\fam\ttfam \twelvett \f@ntkey=7 }\textfont\ttfam=\twelvett
   \h@big=11.9\p@ \h@Big=16.1\p@ \h@bigg=20.3\p@ \h@Bigg=24.5\p@
   \def\caps{\fam\cpfam \twelvecp \f@ntkey=8 }\textfont\cpfam=\twelvecp
   \setbox\strutbox=\hbox{\vrule height 12pt depth 5pt width\z@}%
   \samef@nt}
\def\twelvepoint{\relax
   \textfont0=\twelverm	  \scriptfont0=\ninerm
   \scriptscriptfont0=\sevenrm
   \def\rm{\fam0 \twelverm \f@ntkey=0 }%
   \textfont1=\twelvei		  \scriptfont1=\ninei
   \scriptscriptfont1=\seveni
   \def\oldstyle{\fam1 \twelvei\f@ntkey=1 }%
   \textfont2=\twelvesy	  \scriptfont2=\ninesy
   \scriptscriptfont2=\sevensy
   \textfont3=\twelveex	  \scriptfont3=\twelveex
   \scriptscriptfont3=\twelveex
   \def\it{\fam\itfam \twelveit \f@ntkey=4 }\textfont\itfam=\twelveit
   \def\sl{\fam\slfam \twelvesl \f@ntkey=5 }\textfont\slfam=\twelvesl
   \scriptfont\slfam=\ninesl
   \def\bf{\fam\bffam \twelvebf \f@ntkey=6 }\textfont\bffam=\twelvebf
   \scriptfont\bffam=\ninebf	  \scriptscriptfont\bffam=\sevenbf
   \def\tt{\fam\ttfam \twelvett \f@ntkey=7 }\textfont\ttfam=\twelvett
   \h@big=10.2\p@ \h@Big=13.8\p@ \h@bigg=17.4\p@ \h@Bigg=21.0\p@
   \def\caps{\fam\cpfam \twelvecp \f@ntkey=8 }\textfont\cpfam=\twelvecp
   \setbox\strutbox=\hbox{\vrule height 10pt depth 4pt width\z@}%
   \samef@nt}
\def\tenpoint{\relax
   \textfont0=\tenrm	       \scriptfont0=\sevenrm
   \scriptscriptfont0=\fiverm
   \def\rm{\fam0 \tenrm \f@ntkey=0 }%
   \textfont1=\teni	       \scriptfont1=\seveni
   \scriptscriptfont1=\fivei
   \def\oldstyle{\fam1 \teni \f@ntkey=1 }%
   \textfont2=\tensy	       \scriptfont2=\sevensy
   \scriptscriptfont2=\fivesy
   \textfont3=\tenex	       \scriptfont3=\tenex
   \scriptscriptfont3=\tenex
   \def\it{\fam\itfam \tenit \f@ntkey=4 }\textfont\itfam=\tenit
   \def\sl{\fam\slfam \tensl \f@ntkey=5 }\textfont\slfam=\tensl
   \def\bf{\fam\bffam \tenbf \f@ntkey=6 }\textfont\bffam=\tenbf
   \scriptfont\bffam=\sevenbf	   \scriptscriptfont\bffam=\fivebf
   \def\tt{\fam\ttfam \tentt \f@ntkey=7 }\textfont\ttfam=\tentt
   \def\caps{\fam\cpfam \tencp \f@ntkey=8 }\textfont\cpfam=\tencp
   \h@big=8.5\p@ \h@Big=11.5\p@ \h@bigg=14.5\p@ \h@Bigg=17.5\p@
   \setbox\strutbox=\hbox{\vrule height 8.5pt depth 3.5pt width\z@}%
   \samef@nt}
\newdimen\h@big  \h@big=8.5\p@
\newdimen\h@Big  \h@Big=11.5\p@
\newdimen\h@bigg  \h@bigg=14.5\p@
\newdimen\h@Bigg  \h@Bigg=17.5\p@
\def\big#1{{\hbox{$\left#1\vbox to\h@big{}\right.\n@space$}}}
\def\Big#1{{\hbox{$\left#1\vbox to\h@Big{}\right.\n@space$}}}
\def\bigg#1{{\hbox{$\left#1\vbox to\h@bigg{}\right.\n@space$}}}
\def\Bigg#1{{\hbox{$\left#1\vbox to\h@Bigg{}\right.\n@space$}}}
\normalbaselineskip = 20pt plus 0.2pt minus 0.1pt
\normallineskip = 1.5pt plus 0.1pt minus 0.1pt
\normallineskiplimit = 1.5pt
\newskip\normaldisplayskip
\normaldisplayskip = 20pt plus 5pt minus 10pt
\newskip\normaldispshortskip
\normaldispshortskip = 6pt plus 5pt
\newskip\normalparskip
\normalparskip = 6pt plus 2pt minus 1pt
\newskip\skipregister
\skipregister = 5pt plus 2pt minus 1.5pt
\newif\ifsingl@	   \newif\ifdoubl@
\newif\iftwelv@	   \twelv@true
\def\singlespace{\singl@true\doubl@false\spaces@t}
\def\doublespace{\singl@false\doubl@true\spaces@t}
\def\normalspace{\singl@false\doubl@false\spaces@t}
\def\Tenpoint{\tenpoint\twelv@false\spaces@t}
\def\Twelvepoint{\twelvepoint\twelv@true\spaces@t}
\def\spaces@t{\relax
   \iftwelv@\ifsingl@\subspaces@t3:4;\else\subspaces@t1:1;\fi
   \else\ifsingl@\subspaces@t3:5;\else\subspaces@t4:5;\fi\fi
   \ifdoubl@\multiply\baselineskip by 5 \divide\baselineskip by 4 \fi}
\def\subspaces@t#1:#2;{\baselineskip=\normalbaselineskip
   \multiply\baselineskip by #1\divide\baselineskip by #2%
   \lineskip = \normallineskip
   \multiply\lineskip by #1\divide\lineskip by #2%
   \lineskiplimit = \normallineskiplimit
   \multiply\lineskiplimit by #1\divide\lineskiplimit by #2%
   \parskip = \normalparskip
   \multiply\parskip by #1\divide\parskip by #2%
   \abovedisplayskip = \normaldisplayskip
   \multiply\abovedisplayskip by #1\divide\abovedisplayskip by #2%
   \belowdisplayskip = \abovedisplayskip
   \abovedisplayshortskip = \normaldispshortskip
   \multiply\abovedisplayshortskip by #1%
   \divide\abovedisplayshortskip by #2%
   \belowdisplayshortskip = \abovedisplayshortskip
   \advance\belowdisplayshortskip by \belowdisplayskip
   \divide\belowdisplayshortskip by 2
   \smallskipamount = \skipregister
   \multiply\smallskipamount by #1\divide\smallskipamount by #2%
   \medskipamount = \smallskipamount \multiply\medskipamount by 2
   \bigskipamount = \smallskipamount \multiply\bigskipamount by 4 }
\def\normalbaselines{ \baselineskip=\normalbaselineskip%
   \lineskip=\normallineskip \lineskiplimit=\normallineskip%
   \iftwelv@\else \multiply\baselineskip by 4 \divide\baselineskip by 5%
   \multiply\lineskiplimit by 4 \divide\lineskiplimit by 5%
   \multiply\lineskip by 4 \divide\lineskip by 5 \fi }
\Twelvepoint  % That's the default
\interlinepenalty=50
\interfootnotelinepenalty=5000
\predisplaypenalty=9000
\postdisplaypenalty=500
\hfuzz=1pt
\vfuzz=0.2pt
%
%%%%%%%%%%%%%%%%%%%%%%%%%%%%%%%%%%%%%%%%%%%%%%%%%%%%%%%%%%%%%%%%%%%%%%%%
%
%   Next, I define output routines, footnotes & related stuff.
%
\def\pagecontents{%
   \ifvoid\topins\else\unvbox\topins\vskip\skip\topins\fi
   \dimen@ = \dp255 \unvbox255
   \ifvoid\footins\else\vskip\skip\footins\footrule\unvbox\footins\fi
   \ifr@ggedbottom \kern-\dimen@ \vfil \fi }
\def\makeheadline{\vbox to 0pt{ \skip@=\topskip
   \advance\skip@ by -12pt \advance\skip@ by -2\normalbaselineskip
   \vskip\skip@ \line{\vbox to 12pt{}\the\headline} \vss
   }\nointerlineskip}
\def\makefootline{\baselineskip = 1.5\normalbaselineskip
   \line{\the\footline}}
\newif\iffrontpage
\newif\ifp@genum
\def\nopagenumbers{\p@genumfalse}
\def\pagenumbers{\p@genumtrue}
\pagenumbers
\newtoks\date
\newtoks\Month
\footline={\hss\iffrontpage\else\ifp@genum\tenrm\folio\hss\fi\fi}
\headline={\iffinal\hfil\else\tenrm DRAFT\hfil\the\date\fi}
\def\monthname{\relax\ifcase\month 0/\or January\or February\or
   March\or April\or May\or June\or July\or August\or September\or
   October\or November\or December\else\number\month/\fi}
\date={\monthname\ \number\day, \number\year}
\Month={\monthname\ \number\year}
\countdef\pagenumber=1  \pagenumber=1
\def\advancepageno{\global\advance\pageno by 1
   \ifnum\pagenumber<0 \global\advance\pagenumber by -1
   \else\global\advance\pagenumber by 1 \fi \global\frontpagefalse }
\def\folio{\ifnum\pagenumber<0 \romannumeral-\pagenumber
   \else \number\pagenumber \fi }
\def\footrule{\dimen@=\prevdepth\nointerlineskip
   \vbox to 0pt{\vskip -0.25\baselineskip \hrule width 0.35\hsize \vss}
   \prevdepth=\dimen@ }
\newtoks\foottokens
\foottokens={\Tenpoint\singlespace}
\newdimen\footindent
\footindent=24pt
\def\vfootnote#1{\insert\footins\bgroup  \the\foottokens
   \interlinepenalty=\interfootnotelinepenalty \floatingpenalty=20000
   \splittopskip=\ht\strutbox \boxmaxdepth=\dp\strutbox
   \leftskip=\footindent \rightskip=\z@skip
   \parindent=0.5\footindent \parfillskip=0pt plus 1fil
   \spaceskip=\z@skip \xspaceskip=\z@skip
   \Textindent{$ #1 $}\footstrut\futurelet\next\fo@t}
\def\Textindent#1{\noindent\llap{#1\enspace}\ignorespaces}
\def\footnote#1{\attach{#1}\vfootnote{#1}}

\let\footsymbol=\star
\newcount\lastf@@t	     \lastf@@t=-1
\newcount\footsymbolcount    \footsymbolcount=0
\newif\ifPhysRev
\def\footsymbolgen{\relax \ifPhysRev \iffrontpage \NPsymbolgen\else
   \PRsymbolgen\fi \else \NPsymbolgen\fi
   \global\lastf@@t=\pageno \footsymbol }
\def\NPsymbolgen{\ifnum\footsymbolcount<0 \global\footsymbolcount=0\fi
   {\iffrontpage \else \advance\lastf@@t by 1 \fi
   \ifnum\lastf@@t<\pageno \global\footsymbolcount=0
   \else \global\advance\footsymbolcount by 1 \fi }
   \ifcase\footsymbolcount \fd@f\star\or \fd@f\dagger\or \fd@f\ast\or
   \fd@f\ddagger\or \fd@f\natural\or \fd@f\diamond\or \fd@f\bullet\or
   \fd@f\nabla\else \fd@f\dagger\global\footsymbolcount=0 \fi }
\def\fd@f#1{\xdef\footsymbol{#1}}
\def\PRsymbolgen{\ifnum\footsymbolcount>0 \global\footsymbolcount=0\fi
   \global\advance\footsymbolcount by -1
   \xdef\footsymbol{\sharp\number-\footsymbolcount} }
\def\space@ver#1{\let\@sf=\empty \ifmmode #1\else \ifhmode
   \edef\@sf{\spacefactor=\the\spacefactor}\unskip${}#1$\relax\fi\fi}
\def\attach#1{\space@ver{\strut^{\mkern 2mu #1} }\@sf\ }
%
%%%%%%%%%%%%%%%%%%%%%%%%%%%%%%%%%%%%%%%%%%%%%%%%%%%%%%%%%%%%%%%%%%%%%%%%
%
%   Here come chapter, section, subsection & appendix macros.
%
\newcount\chapternumber	     \chapternumber=0
\newcount\sectionnumber	     \sectionnumber=0
\newcount\equanumber	     \equanumber=0
\let\chapterlabel=\relax
\newtoks\chapterstyle	     \chapterstyle={\Number}
\newskip\chapterskip	     \chapterskip=\bigskipamount
\newskip\sectionskip	     \sectionskip=\medskipamount
\newskip\headskip	     \headskip=8pt plus 3pt minus 3pt
\newdimen\chapterminspace    \chapterminspace=15pc
\newdimen\sectionminspace    \sectionminspace=10pc
\newdimen\referenceminspace  \referenceminspace=25pc
\def\chapterreset{\global\advance\chapternumber by 1
   \ifnum\equanumber<0 \else\global\equanumber=0\fi
   \sectionnumber=0 \makel@bel}
\def\makel@bel{\xdef\chapterlabel{%
   \the\chapterstyle{\the\chapternumber}.}}
\def\sectionlabel{\number\sectionnumber \quad }
\def\alphabetic#1{\count255='140 \advance\count255 by #1\char\count255}
\def\Alphabetic#1{\count255='100 \advance\count255 by #1\char\count255}
\def\Roman#1{\uppercase\expandafter{\romannumeral #1}}
\def\roman#1{\romannumeral #1}
\def\Number#1{\number #1}
\def\unnumberedchapters{\let\makel@bel=\relax \let\chapterlabel=\relax
\let\sectionlabel=\relax \equanumber=-1 }
\def\titlestyle#1{\par\begingroup \interlinepenalty=9999
   \leftskip=0.02\hsize plus 0.23\hsize minus 0.02\hsize
   \rightskip=\leftskip \parfillskip=0pt
   \hyphenpenalty=9000 \exhyphenpenalty=9000
   \tolerance=9999 \pretolerance=9000
   \spaceskip=0.333em \xspaceskip=0.5em
   \iftwelv@\fourteenpoint\else\twelvepoint\fi
   \noindent #1\par\endgroup }
\def\spacecheck#1{\dimen@=\pagegoal\advance\dimen@ by -\pagetotal
   \ifdim\dimen@<#1 \ifdim\dimen@>0pt \vfil\break \fi\fi}
\def\chapter#1{\par \penalty-300 \vskip\chapterskip
   \spacecheck\chapterminspace
   \chapterreset \titlestyle{\chapterlabel \ #1}
   \nobreak\vskip\headskip \penalty 30000
   \wlog{\string\chapter\ \chapterlabel} }

\def\section#1{\par \ifnum\the\lastpenalty=30000\else
   \penalty-200\vskip\sectionskip \spacecheck\sectionminspace\fi
   \wlog{\string\section\ \chapterlabel \the\sectionnumber}
   \global\advance\sectionnumber by 1  \noindent
   {\caps\enspace\chapterlabel \sectionlabel #1}\par
   \nobreak\vskip\headskip \penalty 30000 }
\def\subsection#1{\par
   \ifnum\the\lastpenalty=30000\else \penalty-100\smallskip \fi
   \noindent\undertext{#1}\enspace \vadjust{\penalty5000}}

\def\undertext#1{\vtop{\hbox{#1}\kern 1pt \hrule}}
%

%                             spelling changed (K.Hikasa 4/2/87)
%
\def\APPENDIX#1#2{\par\penalty-300\vskip\chapterskip
   \spacecheck\chapterminspace \chapterreset \xdef\chapterlabel{#1}
   \titlestyle{APPENDIX #2} \nobreak\vskip\headskip \penalty 30000
   \wlog{\string\Appendix\ \chapterlabel} }
\def\Appendix#1{\APPENDIX{#1}{#1}}
\def\appendix{\APPENDIX{A}{}}
%
%%%%%%%%%%%%%%%%%%%%%%%%%%%%%%%%%%%%%%%%%%%%%%%%%%%%%%%%%%%%%%%%%%%%%%%%
%
%   Here come macros for equation numbering.
%
\newif\iffinal \finaltrue
\def\showeqname#1{\iffinal\else\hbox to 0pt{\tentt\kern2mm\string#1\hss}\fi}
\def\showEqname#1{\iffinal\else \hskip 0pt plus 1fill
 \hbox to 0pt{\tentt\kern2mm\string#1\hss}\hskip 0pt plus -1fill\fi}
\def\eqnamedef#1{\relax \ifnum\equanumber<0
   \xdef#1{{\noexpand\rm(\number-\equanumber)}}\global\advance\equanumber by -1
   \else \global\advance\equanumber by 1
   \xdef#1{{\noexpand\rm(\chapterlabel \number\equanumber)}}\fi}
\def\eqnamenewdef#1#2{\relax \ifnum\equanumber<0
   \xdef#1{{\noexpand\rm(\number-\equanumber#2)}}\global\advance\equanumber
   by -1 \else \global\advance\equanumber by 1
   \xdef#1{{\noexpand\rm(\chapterlabel \number\equanumber#2)}}\fi}
\def\eqnameolddef#1#2{\relax \ifnum\equanumber<0
   \global\advance\equanumber by 1
   \xdef#1{{\noexpand\rm(\number-\equanumber#2)}}\global\advance\equanumber
   by -1 \else \xdef#1{{\noexpand\rm(\chapterlabel \number\equanumber#2)}}\fi}
\def\eqname#1{\eqnamedef{#1}#1}
\def\eqnamenew#1#2{\eqnamenewdef{#1}{#2}#1}
\def\eqnameold#1#2{\eqnameolddef{#1}{#2}#1}
\def\eq{\eqname\lasteq}
\def\eqa{\eqnamenew\lasteq a}
\def\eqb{\eqnameold\lasteq b}
\def\eqc{\eqnameold\lasteq c}
\def\eqd{\eqnameold\lasteq d}
\def\eqnew#1{\eqnamenew\lasteq{#1}}
\def\eqold#1{\eqnameold\lasteq{#1}}
\def\eq@@{\ifinner\let\eqn@=\relax\else\let\eqn@=\eqno\fi\eqn@}
\def\Eq{\eq@@\eq}
\def\Eqnew#1{\eq@@\eqnew{#1}}
\def\Eqold#1{\eq@@\eqold{#1}}
\def\Eqa{\eq@@\eqa}
\def\Eqb{\eq@@\eqb}
\def\Eqc{\eq@@\eqc}
\def\Eqd{\eq@@\eqd}
\def\Eqn#1{\eq@@\eqname{#1}\showeqname{#1}}
\def\Eqnnew#1#2{\eq@@\eqnamenew{#2}{#1}\showeqname{#1}}
\def\Eqnold#1#2{\eq@@\eqnameold{#2}{#1}\showeqname{#1}}
\def\Eqna#1{\eq@@\eqnamenew{#1}a\showeqname{#1}}
\def\Eqnb#1{\eq@@\eqnameold{#1}b\showeqname{#1}}
\def\Eqnc#1{\eq@@\eqnameold{#1}c\showeqname{#1}}
\def\Eqnd#1{\eq@@\eqnameold{#1}d\showeqname{#1}}

%

%
%%%%%%%%%%%%%%%%%%%%%%%%%%%%%%%%%%%%%%%%%%%%%%%%%%%%%%%%%%%%%%%%%%%%%%%%
%   Here come items and lists
%
\def\GENITEM#1;#2{\par \hangafter=0 \hangindent=#1
   \Textindent{$ #2 $}\ignorespaces}
\outer\def\newitem#1=#2;{\gdef#1{\GENITEM #2;}}
\newdimen\itemsize		  \itemsize=30pt
\newitem\item=1\itemsize;
\newitem\sitem=1.75\itemsize;	  
\newitem\ssitem=2.5\itemsize;	  
\outer\def\newlist#1=#2&#3&#4;{\toks0={#2}\toks1={#3}%
   \count255=\escapechar \escapechar=-1
   \alloc@0\list\countdef\insc@unt\listcount	 \listcount=0
   \edef#1{\par
      \countdef\listcount=\the\allocationnumber
      \advance\listcount by 1
      \hangafter=0 \hangindent=#4
      \Textindent{\the\toks0{\listcount}\the\toks1}}
   \expandafter\expandafter\expandafter
   \edef\c@t#1{begin}{\par
      \countdef\listcount=\the\allocationnumber \listcount=1
      \hangafter=0 \hangindent=#4
      \Textindent{\the\toks0{\listcount}\the\toks1}}
   \expandafter\expandafter\expandafter
   \edef\c@t#1{con}{\par \hangafter=0 \hangindent=#4 \noindent}
   \escapechar=\count255}
\def\c@t#1#2{\csname\string#1#2\endcsname}
\newlist\point=\Number&.&1.0\itemsize;
\newlist\subpoint=(\alphabetic&)&1.75\itemsize;
\newlist\subsubpoint=(\roman&)&2.5\itemsize;
%

%
%%%%%%%%%%%%%%%%%%%%%%%%%%%%%%%%%%%%%%%%%%%%%%%%%%%%%%%%%%%%%%%%%%%%%%%%
%
%   Here come macros for references, figures & tables.
%
\def\keepspacefactor{\let\@sf=\empty \ifhmode
   \edef\@sf{\spacefactor=\the\spacefactor\relax}\relax\fi}
\newcount\footcount \footcount=0
\def\Footnote{\global\advance\footcount by 1 \footnote{\the\footcount}}
\def\footnote#1{\keepspacefactor\refattach{#1}\vfootnote{#1}}

\def\nonfrenchspacing{\sfcode\lq\.=3000 \sfcode\lq\?=3001 \sfcode\lq\!=3001
 \sfcode\lq\:=2000 \sfcode\lq\;=1500 \sfcode\lq\,=1250 }

\nonfrenchspacing
%%%%%%%%%%%%%%%%%%%%%%%%%%
\newcount\referencecount     \referencecount=0
\newif\ifreferenceopen	     \newwrite\referencewrite
\newtoks\rw@toks
\newcount\lastrefsbegincount \lastrefsbegincount=0
\def\refsend{\refmark{\count255=\referencecount
   \advance\count255 by-\lastrefsbegincount
   \ifcase\count255 \number\referencecount
   \or \number\lastrefsbegincount,\number\referencecount
   \else \number\lastrefsbegincount-\number\referencecount \fi}}
\def\refch@ck{\chardef\rw@write=\referencewrite
   \ifreferenceopen \else \referenceopentrue
   \immediate\openout\referencewrite=reference.aux \fi}
%
% In \obeyendofline, we say `\let^^M=\relax
{\catcode`\^^M=\active % these lines must end with %
  \gdef\obeyendofline{\catcode`\^^M\active \let^^M\ }}%
%
% In \ignoreendofline, we say `\let^^M=\relax
{\catcode`\^^M=\active % these lines must end with %
  \gdef\ignoreendofline{\catcode`\^^M=5}}
{\obeyendofline\gdef\rw@start#1{\def\t@st{#1}\ifx\t@st\blankend%
\endgroup {\@sf} \relax \else \ifx\t@st\bl@nkend \endgroup {\@sf} \relax%
\else \rw@begin#1
\backtotext
\fi \fi } }
{\obeyendofline\gdef\rw@begin#1
{\def\n@xt{#1}\rw@toks={#1}\relax%
\rw@next}}
\def\blankend{}
{\obeylines\gdef\bl@nkend{
}}
\newif\iffirstrefline  \firstreflinetrue
\def\rwr@teswitch{\ifx\n@xt\blankend \let\n@xt=\rw@begin %
 \else\iffirstrefline \global\firstreflinefalse%
\immediate\write\rw@write{\noexpand\obeyendofline \the\rw@toks}%
\let\n@xt=\rw@begin%
      \else\ifx\n@xt\rw@@d \def\n@xt{\immediate\write\rw@write{%
	\noexpand\ignoreendofline}\endgroup \@sf}%
	     \else \immediate\write\rw@write{\the\rw@toks}%
	     \let\n@xt=\rw@begin\fi\fi \fi}
\def\rw@next{\rwr@teswitch\n@xt}
\def\rw@@d{\backtotext} \let\rw@end=\relax
\let\backtotext=\relax

%%%%%%%%%%%%%%%%%%%%%%%%%%%%%%%
\newdimen\refindent	\refindent=20pt
\newmuskip\refskip
\newmuskip\regularrefskip \regularrefskip=2mu
\newmuskip\specialrefskip \specialrefskip=-2mu
\def\refattach#1{\@sf \ifhmode\ifnum\spacefactor=1250 \refskip=\specialrefskip
 \else\ifnum\spacefactor=3000 \refskip=\specialrefskip
 \else\ifnum\spacefactor=1001 \refskip=\specialrefskip
 \else \refskip=\regularrefskip \fi\fi\fi
 \else \refskip=\regularrefskip \fi
 \ref@ttach{\strut^{\mkern\refskip #1}}}
\def\ref@ttach#1{\ifmmode #1\else\ifhmode\unskip${}#1$\relax\fi\fi{\@sf}}
\def\PLrefmark#1{ [#1]{\@sf}}
\def\NPrefmark#1{\refattach{\scriptstyle [ #1 ] }}
\let\PRrefmark=\refattach
\def\refmark{\keepspacefactor\refm@rk}
\def\refm@rk#1{\relax\therefm@rk{#1}}
\def\originalrefs{\let\therefm@rk=\NPrefmark}
\def\PRrefs{\let\therefm@rk=\PRrefmark \let\therefitem=\PRrefitem}
\def\PLrefs{\let\therefm@rk=\PLrefmark \let\therefitem=\PLrefitem}
\def\PRrefitem#1{\refitem{#1.}}
\def\PLrefitem#1{\refitem{[#1]}}
\let\therefitem=\PRrefitem
\def\refitem#1{\par \hangafter=0 \hangindent=\refindent \Textindent{#1}}
\def\REFNUM#1{\eatspace\keepspacefactor\refch@ck \firstreflinetrue%
 \global\advance\referencecount by 1 \xdef#1{\the\referencecount}}
\def\eatspace{\ifhmode\unskip\fi}
\def\refnum#1{\keepspacefactor\refch@ck \firstreflinetrue%
 \global\advance\referencecount by 1 \xdef#1{\the\referencecount}\refend}
\def\REF#1{\REFNUM#1%
 \immediate\write\referencewrite{%
 \noexpand\therefitem{#1}}%
\begingroup\obeyendofline\rw@start}
\def\ref{\refnum\?%
 \immediate\write\referencewrite{\noexpand\therefitem{\?}}%
\begingroup\obeyendofline\rw@start}
\def\Ref#1{\refnum#1%
 \immediate\write\referencewrite{\noexpand\therefitem{#1}}%
\begingroup\obeyendofline\rw@start}
\def\REFS#1{\REFNUM#1\global\lastrefsbegincount=\referencecount
\immediate\write\referencewrite{\noexpand\therefitem{#1}}%
\begingroup\obeyendofline\rw@start}
\def\refend{\refm@rk{\number\referencecount}}
%

%

%%%%%%%%%%%%%%%%%%%%%%%%
\newcount\figurecount	  \figurecount=0
\newif\iffigureopen	  \newwrite\figurewrite
\newdimen\digitwidth \setbox0=\hbox{\rm0} \digitwidth=\wd0
\def\zerophant{\kern\digitwidth}
\def\FIGNUM#1{\keepspacefactor\figch@ck \firstreflinetrue%
\global\advance\figurecount by 1 \xdef#1{\the\figurecount}}
\def\figch@ck{\chardef\rw@write=\figurewrite \iffigureopen\else
   \immediate\openout\figurewrite=figures.aux
   \figureopentrue\fi}
\def\figitem#1{\par\indent \hangindent2\parindent \textindent{Fig. #1\ }}
\def\FIGLABEL#1{\ifnum\number#1<10 \def\figlabel{#1.\zerophant}\else%
\def\figlabel{#1.}\fi}
\def\FIG#1{\FIGNUM#1\FIGLABEL#1%
\immediate\write\figurewrite{\noexpand\figitem{\figlabel}}%
\begingroup\obeyendofline\rw@start}
\def\Figname#1{\FIGNUM#1Fig.~#1\FIGLABEL#1%
\immediate\write\figurewrite{\noexpand\figitem{\figlabel}}%
\begingroup\obeyendofline\rw@start}
\def\fig{\FIGNUM\? fig.~\? \FIGLABEL\?
\immediate\write\figurewrite{\noexpand\figitem{\figlabel}}%
\begingroup\obeyendofline\rw@start}
\def\figure{\FIGNUM\? figure~\? \FIGLABEL\?
\immediate\write\figurewrite{\noexpand\figitem{\figlabel}}%
\begingroup\obeyendofline\rw@start}
\def\Fig{\FIGNUM\? Fig.~\? \FIGLABEL\?
\immediate\write\figurewrite{\noexpand\figitem{\figlabel}}%
\begingroup\obeyendofline\rw@start}
\def\Figure{\FIGNUM\? Figure~\? \FIGLABEL\?
\immediate\write\figurewrite{\noexpand\figitem{\figlabel}}%
\begingroup\obeyendofline\rw@start}
\def\figout{\par \penalty-400 \vskip\chapterskip
  \spacecheck\referenceminspace \immediate\closeout\figurewrite
  \figureopenfalse
  \line{\fourteenrm
   \hfil FIGURE CAPTION\ifnum\figurecount=1 \else S \fi\hfil}
  \vskip\headskip
  \input figures.aux
  }
%%%%%%%%%%%%%%%%%%%%%
\newcount\tablecount	 \tablecount=0
\newif\iftableopen	 \newwrite\tablewrite
\def\tabch@ck{\chardef\rw@write=\tablewrite \iftableopen\else
   \immediate\openout\tablewrite=tables.aux
   \tableopentrue\fi}
\def\TABNUM#1{\keepspacefactor\tabch@ck \firstreflinetrue%
\global\advance\tablecount by 1 \xdef#1{\the\tablecount}}
\def\tableitem#1{\par\indent \hangindent2\parindent \textindent{Table #1\ }}
\def\TABLE#1{\TABNUM#1\FIGLABEL#1%
\immediate\write\tablewrite{\noexpand\tableitem{\figlabel}}%
\begingroup\obeyendofline\rw@start}
\def\Table{\TABNUM\? Table~\?\FIGLABEL\?%
\immediate\write\tablewrite{\noexpand\tableitem{\figlabel}}%
\begingroup\obeyendofline\rw@start}
\def\tabout{\par \penalty-400 \vskip\chapterskip
  \spacecheck\referenceminspace \immediate\closeout\tablewrite \tableopenfalse
  \line{\fourteenrm\hfil TABLE CAPTION\ifnum\tablecount=1 \else S\fi\hfil}
  \vskip\headskip
  \input tables.aux
  }
\PRrefs
\def\etal{{\it et al.}}
%
%%%%%%%%%%%%%%%%%%%%%%%%%%%%%%%%%%%%%%%%%%%%%%%%%%%%%%%%%%%%%%%%%%%%%%%%
\def\masterreset{\global\pagenumber=1 \global\chapternumber=0
   \global\equanumber=0 \global\sectionnumber=0
   \global\referencecount=0 \global\figurecount=0 \global\tablecount=0 }
\def\FRONTPAGE{\ifvoid255\else\vfill\penalty-2000\fi
      \masterreset\global\frontpagetrue
      \global\lastf@@t=0 \global\footsymbolcount=0}

\def\papersize{\hsize=35pc\vsize=50pc\hoffset=1pc\voffset=6pc
  \skip\footins=\bigskipamount}
\def\paperstyle{\normalspace\papersize}
\paperstyle
%%%%%%%%%%%%%%%%%%%%%%%%%%%%%%%%%%%%%%%%%%%%%%%%%%%%%%%%%%%%%%%%%%%%%%%
%   Here come macros for title pages.
%%%%%%%%%%%%%%%%%%%%%%%%%%%%%%%%%%%%%%%%%%%%%%%%%%%%%%%%%%%%%%%%%%%%%%
\newskip\frontpageskip
\newtoks\Pubnum \newtoks\pubnum
\newtoks\s@condpubnum \newtoks\th@rdpubnum
\newif\ifs@cond \s@condfalse
\newif\ifth@rd \th@rdfalse
\newif\ifp@bblock  \p@bblocktrue
\newcount\Year
\def\Yearset{\Year=\year \advance\Year by -1900
 \ifnum\month<4 \advance\Year by -1 \fi}
\def\PH@SR@V{\doubl@true \baselineskip=24.1pt plus 0.2pt minus 0.1pt
	     \parskip= 3pt plus 2pt minus 1pt }
\def\PHYSREV{\paperstyle\PhysRevtrue\PH@SR@V}
\def\titlepage{\Yearset\FRONTPAGE\paperstyle\ifPhysRev\PH@SR@V\fi
   \ifp@bblock\p@bblock\fi}
\def\nopubblock{\p@bblockfalse}
\def\endpage{\vfil\break}
\frontpageskip=1\medskipamount plus .5fil
\Pubnum={TU--\the\pubnum }
\pubnum={ }
\def\secondpubnum#1{\s@condtrue\s@condpubnum={#1}}
\def\thirdpubnum#1{\th@rdtrue\th@rdpubnum={#1}}
\def\p@bblock{\begingroup \tabskip=\hsize minus \hsize
   \baselineskip=1.5\ht\strutbox \topspace-2\baselineskip
   \halign to\hsize{\strut ##\hfil\tabskip=0pt\crcr
   \the\Pubnum\cr
   \ifs@cond \the\s@condpubnum\cr\fi
   \ifth@rd \the\th@rdpubnum\cr\fi
   \the\Month \cr}\endgroup}
\def\title#1{\hrule height0pt depth0pt
   \vskip\frontpageskip \titlestyle{#1} \vskip\headskip }
\def\author#1{\vskip\frontpageskip\titlestyle{\twelvecp #1}\nobreak}

\def\address#1{\par\kern 5pt\titlestyle{\twelvepoint\it #1}}
\def\andaddress{\par\kern 5pt \centerline{\sl and} \address}
\def\abstract{\vskip\frontpageskip\centerline{\fourteenrm ABSTRACT}
 \vskip\headskip }

%
%
%%%%%%%%%%%%%%%%%%%%%%%%%%%%%%%%%%%%%%%%%%%%%%%%%%%%%%%%%%%%%%%%%%%%%%%%
%   Miscellaneous macros
%

\def\\{\relax\ifmmode\backslash\else$\backslash$\fi}
\def\globaleqnumbers{\relax\if\equanumber<0\else\global\equanumber=-1\fi}
\def\nextline{\unskip\nobreak\hskip\parfillskip\break}

\def\topspace{\hrule height 0pt depth 0pt \vskip}

\def\VEV#1{\left\langle #1\right\rangle}

\let\int=\intop 
\def\prop{\mathrel{{\mathchoice{\pr@p\scriptstyle}{\pr@p\scriptstyle}%
 {\pr@p\scriptscriptstyle}{\pr@p\scriptscriptstyle} }}}
\def\pr@p#1{\setbox0=\hbox{$\cal #1 \char'103$}
   \hbox{$\cal #1 \char'117$\kern-.4\wd0\box0}}
\def\lsim{\mathrel{\mathpalette\@versim<}}
\def\gsim{\mathrel{\mathpalette\@versim>}}
\def\@versim#1#2{\lower0.2ex\vbox{\baselineskip\z@skip\lineskip\z@skip
  \lineskiplimit\z@\ialign{$\m@th#1\hfil##\hfil$\crcr#2\crcr\sim\crcr}}}
%
% % % % % % % % % % % % % % % % % % % % % % % % % % % % % % % % % % % %
%
%   Finally, some bug fixings.
%
\let\sec@nt=\sec
\def\sec{\relax\ifmmode\let\n@xt=\sec@nt\else\let\n@xt\section\fi\n@xt}
\def\obsolete#1{\message{Macro \string #1 is obsolete.}}
\def\firstsec#1{\obsolete\firstsec \section{#1}}
\def\firstsubsec#1{\obsolete\firstsubsec \subsection{#1}}
\def\thispage#1{\obsolete\thispage \global\pagenumber=#1\frontpagefalse}
\def\thischapter#1{\obsolete\thischapter \global\chapternumber=#1}
\def\nextequation#1{\obsolete\nextequation \global\equanumber=#1
   \ifnum\the\equanumber>0 \global\advance\equanumber by 1 \fi}
\def\BOXITEM{\afterassigment\B@XITEM\setbox0=}
\def\B@XITEM{\par\hangindent\wd0 \noindent\box0 }
%

%%%%%%%%%%%%%%%%%%%%%%%%%%%%%%%%%%%%%%%%%%%%%%%%%%%%%%%%%%%%%%%%%%%%%%%%
%   That's about it
%
\catcode`\@=12
%
%\message{Done }
%\everyjob{\input myphyx }
%%%%%%%%%%%%%%%%%%%%%%%%%%%%%%%%%%%%%%%%%%%%%%%%%%%%%%%%%%%%%%%%%%%%%%%%%%%%%
\let\to=\rightarrow
\foottokens={\vskip 1\jot\Tenpoint }

\def\r{\noalign{\vskip 3\jot }}

\def\tfrac#1#2{{\textstyle{#1\over #2}}}

\def\({\hbox{\sevenrm(}\mskip-1mu}
\def\){\hbox{\sevenrm)}}

\def\tsum{\mathop{\textstyle\sum}\limits}

\mathchardef\REAL="023C
\mathchardef\IMAG="023D

\def\hidehrule#1#2{\kern-#1 \hrule height#1 depth#2 \kern-#2 }
\def\hidevrule#1#2{\kern-#1{\dimen0=#1 \advance\dimen0 by #2%
\vrule width\dimen0}\kern-#2 }
\def\makeblankbox#1#2{\hbox{\lower\dp0\vbox{\hidehrule{#1}{#2}%
\kern-#1 \hbox to \wd0{\hidevrule{#1}{#2}\raise\ht0\vbox to #1{}%
\lower\dp0\vtop to #1{}\hfil\hidevrule{#2}{#1}}\kern -#1\hidehrule{#2}{#1}}}}
\def\dAlembert{\setbox0=\hbox{$\Sigma$}\kern .1em \lower .1ex
\makeblankbox{.25pt}{.25pt}\kern .1em }

{\catcode`\|=0 \catcode`\\=12 % | is temporary escape character
  |obeylines|gdef|doverbatim^^M#1\endverbatim{#1|endgroup}}
%\newcount\lineno % the number of file lines listed

% \everypar{\advance\lineno by 1 \llap{\sevenrm\the\lineno\ \ }}}
 {\obeyspaces\global\let =\ } % let active space = control space

\def\goodmood{\hbox{$\rlap{\kern.35ex$\scriptscriptstyle\smile
$}\rlap{\kern.55ex\lower.45ex\hbox{$\mathchar"707F$}}\bigcirc$}}
\def\badmood{\hbox{$\rlap{\kern.35ex$\scriptscriptstyle\frown
$}\rlap{\kern.55ex\lower.45ex\hbox{$\mathchar"707F$}}\bigcirc$}}
\def\verybadmood{\hbox{$\rlap{\kern.5ex\lower1.25ex\hbox{$\mathchar"0365
$}}\rlap{\kern.55ex\lower.45ex\hbox{$\mathchar"707F$}}\bigcirc$}}
\def\leftmood{\hbox{$\rlap{\kern.5ex\lower1.25ex\hbox{$\mathchar"0365
$}}\rlap{\kern.35ex\lower.45ex\hbox{$\mathchar"707F$}}\bigcirc$}}
\def\rightmood{\hbox{$\rlap{\kern.5ex\lower1.25ex\hbox{$\mathchar"0365
$}}\rlap{\kern.75ex\lower.45ex\hbox{$\mathchar"707F$}}\bigcirc$}}
%%%%%%%%%%%%%%%%%%
%
\def\GENITEM#1;#2{\par \hangafter=0 \hangindent=#1
    \Textindent{#2}\ignorespaces}
%
%\outer\def\newitem#1=#2;{\gdef#1{\GENITEM #2;}}
%
\newitem\item=1\itemsize;
\newitem\sitem=1.75\itemsize;	  
\newitem\ssitem=2.5\itemsize;	  
\newitem\appitem=2.9\itemsize;
%
%%%%%%%%%%%%%%%%%%%
\font\seventeenbi=cmmib10 scaled\magstep3
\font\twelvebi=cmmib10 scaled\magstep1
\font\twelvebsy=cmbsy10 scaled\magstep1
\def\title#1{\hrule height0pt depth0pt
   \vskip\frontpageskip \titlestyle{\baselineskip=35pt \begingroup
   \textfont1=\seventeenbi \scriptfont1=\twelvebi
   \scriptfont0=\twelvebf \scriptfont2=\twelvebsy
   \seventeenbf #1 \endgroup} \vskip\headskip }
\def\author#1{\vskip\frontpageskip\titlestyle{\fourteenrm #1}\nobreak}
\def\eV{\hbox{$\,\rm eV$}} 
\def\MeV{\hbox{$\,\rm MeV$}} \def\GeV{\hbox{$\,\rm GeV$}}

 \def\m{\hbox{$\,\rm m$}}
\def\cm{\hbox{$\,\rm cm$}}

\def\jnfont{\rm}
\def\APPB#1,{{\jnfont Acta Phys.\ Polon.}\ {\bf B#1},}
\def\AP#1,{{\jnfont Ann.\ Phys.\ (N.Y.)} {\bf #1},}
\def\ApJ#1,{{\jnfont Astrophys.\ J.}\ {\bf #1},}
\def\EpL#1,{{\jnfont Europhys.\ Lett.}\ {\bf #1},}
\def\IJMPA#1,{{\jnfont Int.\ J.\ Mod.\ Phys.\ A}~{\bf #1},}
\def\JETP#1,{{\jnfont Sov.\ Phys.\ JETP}\ {\bf #1},}
\def\JETPL#1,{{\jnfont JETP Lett.}\ {\bf #1},}
\def\JMP#1,{{\jnfont J. Math.\ Phys.}\ {\bf #1},}
\def\LNC#1,{{\jnfont Lett.\ Nuovo Cimento} {\bf #1},}
\def\MPLA#1,{{\jnfont Mod.\ Phys.\ Lett.\ A}~{\bf #1},}
\def\NC#1,{{\jnfont Nuovo Cimento} {\bf #1},}
\def\NP#1,{{\jnfont Nucl.\ Phys.}\ {\bf #1},}
\def\NPA#1,{{\jnfont Nucl.\ Phys.}\ {\bf A#1},}
\def\NPB#1,{{\jnfont Nucl.\ Phys.}\ {\bf B#1},}
\def\Physica#1,{{\jnfont Physica}\ {\bf #1},}
\def\PL#1,{{\jnfont Phys.\ Lett.}\ {\bf #1},}
\def\PLB#1,{{\jnfont Phys.\ Lett.\ B}~{\bf #1},}
\def\PRep#1,{{\jnfont Phys.\ Rep.}\ {\bf #1},}
\def\PR#1,{{\jnfont Phys.\ Rev.}\ {\bf #1},}
\def\PRD#1,{{\jnfont Phys.\ Rev.\ D}~{\bf #1},}
\def\PRL#1,{{\jnfont Phys.\ Rev.\ Lett.}\ {\bf #1},}
\def\PTP#1,{{\jnfont Prog.\ Theor.\ Phys.}\ {\bf #1},}
\def\PZETF#1,{{\jnfont Pis'ma Zh.\ Eksp.\ Teor.\ Fiz.}\ {\bf #1},}
\def\RMP#1,{{\jnfont Rev.\ Mod.\ Phys.}\ {\bf #1},}
\def\SJNP#1,{{\jnfont Sov.\ J. Nucl.\ Phys.}\ {\bf #1},}
\def\YaF#1,{{\jnfont Yad.\ Fiz.}\ {\bf #1},}
\def\ZETF#1,{{\jnfont Zh.\ Eksp.\ Teor.\ Fiz.}\ {\bf #1},}
\def\ZPC#1,{{\jnfont Z. Phys.\ C} {\bf #1},}
%%%%%%%%%%%%%%%%%%%%%%%%%%%%%%%%%%%%%%%%%%%%%%%%%%%%%%%%%%%%%%%%%%%%
% boldfont.tex
% written by K. Hikasa
% revised May 14, 1990
% Rev.    Dec.25, 1991
%%%%%%%%%%%%%%%%%%%%%%%%%%%%%%%%%%%%%%%%%%%%%%%%%%%%%%%%%%%%%%%%%%%%
\font\fourteenbi=cmmib10 scaled\magstep2   \skewchar\fourteenbi='177
\font\twelvebi=cmmib10 scaled\magstep1     \skewchar\twelvebi='177
\font\elevenbi=cmmib10 scaled\magstephalf  \skewchar\elevenbi='177
\font\tenbi=cmmib10                        \skewchar\tenbi='177
\font\fourteenbsy=cmbsy10 scaled\magstep2  \skewchar\fourteenbsy='60
\font\twelvebsy=cmbsy10 scaled\magstep1    \skewchar\twelvebsy='60
\font\elevenbsy=cmbsy10 scaled\magstephalf \skewchar\elevenbsy='60
\font\tenbsy=cmbsy10                       \skewchar\tenbsy='60
\font\fourteenbsl=cmbxsl10 scaled\magstep2
\font\twelvebsl=cmbxsl10 scaled\magstep1
\font\elevenbsl=cmbxsl10 scaled\magstephalf
\font\tenbsl=cmbxsl10
\font\fourteenbit=cmbxti10 scaled\magstep2
\font\twelvebit=cmbxti10 scaled\magstep1
\font\elevenbit=cmbxti10 scaled\magstephalf
\font\tenbit=cmbxti10
\catcode\lq\@=11
\def\fourteenbold{\relax
    \textfont0=\fourteenbf	    \scriptfont0=\tenbf
    \scriptscriptfont0=\sevenbf
     \def\rm{\fam0 \fourteenbf \f@ntkey=0 }\relax
    \textfont1=\fourteenbi	    \scriptfont1=\tenbi
    \scriptscriptfont1=\seveni
     \def\oldstyle{\fam1 \fourteenbi\f@ntkey=1 }\relax
    \textfont2=\fourteenbsy	    \scriptfont2=\tenbsy
    \scriptscriptfont2=\sevensy
    \textfont3=\fourteenex     \scriptfont3=\fourteenex
    \scriptscriptfont3=\fourteenex
    \def\it{\fam\itfam \fourteenbit\f@ntkey=4 }\textfont\itfam=\fourteenbit
    \def\sl{\fam\slfam \fourteenbsl\f@ntkey=5 }\textfont\slfam=\fourteenbsl
    \scriptfont\slfam=\tensl
    \def\bf{\fam\bffam \fourteenrm\f@ntkey=6 }\textfont\bffam=\fourteenrm
    \scriptfont\bffam=\tenrm	 \scriptscriptfont\bffam=\sevenrm
    \def\tt{\fam\ttfam \twelvett \f@ntkey=7 }\textfont\ttfam=\twelvett
    \h@big=11.9\p@ \h@Big=16.1\p@ \h@bigg=20.3\p@ \h@Bigg=24.5\p@
    \def\caps{\fam\cpfam \twelvecp \f@ntkey=8 }\textfont\cpfam=\twelvecp
    \setbox\strutbox=\hbox{\vrule height 12pt depth 5pt width\z@}\relax
    \samef@nt}
\def\twelvebold{\relax
    \textfont0=\twelvebf	  \scriptfont0=\ninebf
    \scriptscriptfont0=\sevenbf
     \def\rm{\fam0 \twelvebf \f@ntkey=0 }\relax
    \textfont1=\twelvebi	  \scriptfont1=\ninei
    \scriptscriptfont1=\seveni
     \def\oldstyle{\fam1 \twelvebi\f@ntkey=1 }\relax
    \textfont2=\twelvebsy	  \scriptfont2=\ninesy
    \scriptscriptfont2=\sevensy
    \textfont3=\twelveex	  \scriptfont3=\twelveex
    \scriptscriptfont3=\twelveex
    \def\it{\fam\itfam \twelvebit \f@ntkey=4 }\textfont\itfam=\twelvebit
    \def\sl{\fam\slfam \twelvebsl \f@ntkey=5 }\textfont\slfam=\twelvebsl
    \scriptfont\slfam=\ninesl
    \def\bf{\fam\bffam \twelverm \f@ntkey=6 }\textfont\bffam=\twelverm
    \scriptfont\bffam=\ninerm	  \scriptscriptfont\bffam=\sevenrm
    \def\tt{\fam\ttfam \twelvett \f@ntkey=7 }\textfont\ttfam=\twelvett
    \h@big=10.2\p@ \h@Big=13.8\p@ \h@bigg=17.4\p@ \h@Bigg=21.0\p@
    \def\caps{\fam\cpfam \twelvecp \f@ntkey=8 }\textfont\cpfam=\twelvecp
    \setbox\strutbox=\hbox{\vrule height 10pt depth 4pt width\z@}\relax
    \samef@nt}
\def\elevenbold{\relax
    \textfont0=\elevenbf	  \scriptfont0=\ninebf
    \scriptscriptfont0=\sixbf
     \def\rm{\fam0 \elevenbf \f@ntkey=0 }\relax
    \textfont1=\elevenbi	  \scriptfont1=\ninei
    \scriptscriptfont1=\sixi
     \def\oldstyle{\fam1 \elevenbi\f@ntkey=1 }\relax
    \textfont2=\elevenbsy	  \scriptfont2=\ninesy
    \scriptscriptfont2=\sixsy
    \textfont3=\elevenex	  \scriptfont3=\elevenex
    \scriptscriptfont3=\elevenex
    \def\it{\fam\itfam \elevenbit \f@ntkey=4 }\textfont\itfam=\elevenbit
    \def\sl{\fam\slfam \elevenbsl \f@ntkey=5 }\textfont\slfam=\elevenbsl
    \scriptfont\slfam=\ninesl
    \def\bf{\fam\bffam \elevenrm \f@ntkey=6 }\textfont\bffam=\elevenrm
    \scriptfont\bffam=\ninerm	  \scriptscriptfont\bffam=\sixrm
    \def\tt{\fam\ttfam \eleventt \f@ntkey=7 }\textfont\ttfam=\eleventt
    \h@big=9.311\p@ \h@Big=12.6\p@ \h@bigg=15.88\p@ \h@Bigg=19.17\p@
    \def\caps{\fam\cpfam \elevencp \f@ntkey=8 }\textfont\cpfam=\elevencp
    \setbox\strutbox=\hbox{\vrule height 9pt depth 4pt width\z@}\relax
    \samef@nt}
\def\tenbold{\relax
    \textfont0=\tenbf	       \scriptfont0=\sevenrm
    \scriptscriptfont0=\fiverm
    \def\rm{\fam0 \tenrm \f@ntkey=0 }\relax
    \textfont1=\tenbi	       \scriptfont1=\seveni
    \scriptscriptfont1=\fivei
    \def\oldstyle{\fam1 \tenbi \f@ntkey=1 }\relax
    \textfont2=\tenbsy	       \scriptfont2=\sevensy
    \scriptscriptfont2=\fivesy
    \textfont3=\tenex	       \scriptfont3=\tenex
    \scriptscriptfont3=\tenex
    \def\it{\fam\itfam \tenbit \f@ntkey=4 }\textfont\itfam=\tenbit
    \def\sl{\fam\slfam \tenbsl \f@ntkey=5 }\textfont\slfam=\tenbsl
    \def\bf{\fam\bffam \tenrm \f@ntkey=6 }\textfont\bffam=\tenrm
    \scriptfont\bffam=\sevenrm  \scriptscriptfont\bffam=\fiverm
    \def\tt{\fam\ttfam \tentt \f@ntkey=7 }\textfont\ttfam=\tentt
    \def\caps{\fam\cpfam \tencp \f@ntkey=8 }\textfont\cpfam=\tencp
    \h@big=8.5\p@ \h@Big=11.5\p@ \h@bigg=14.5\p@ \h@Bigg=17.5\p@
    \setbox\strutbox=\hbox{\vrule height 8.5pt depth 3.5pt width\z@}\relax
    \samef@nt}
\def\bold{\iftwelv@\twelvebold\else\ifelev@n\elevenbold\else\tenbold\fi\fi}
\catcode\lq\@=12
%%%%%%%%%%%%%%%%%%%%%%%%%
\font\seventeenbi=cmmib10 scaled\magstep3
\advance\vsize by 36pt
\advance\voffset by -36pt
\PLrefs
\def\half{\tfrac12}
\def\GF{G_{\!F}}
\def\mH{m_H}
\def\SU{{\rm SU}}
\catcode`\@=13 \def@{{\vphantom.}}
%%%%%%%%%%%%%%%%%%%%%%%%%%%%%%%%%%%%%%%%%%%%%%%%%%%%%%%%%%%%%%%%%%%%%%
\newcount\footnotecount \footnotecount=0
\let\ofootnote=\footnote
\def\footnote{\global\advance\footnotecount by 1
 \ofootnote{\ddagger\the\footnotecount}}
%%%%%%%%%%%%%%%%%%%%%%%%%%%%%%%%%%%%%%%%%%%%%%%%%%%%%%%%%%%%%%%%%%%%%%
\def\chapter#1{\par \penalty-300 \vskip\chapterskip
   \spacecheck\chapterminspace
   \chapterreset \par\noindent{\twelvebold\chapterlabel \ #1}\par
   \nobreak\vskip\headskip \penalty 30000
   \wlog{\string\chapter\ \chapterlabel} }
\chapterminspace=120pt

\def\figout{\par \penalty-400 \vskip\chapterskip
   \spacecheck\referenceminspace \immediate\closeout\figurewrite
   \figureopenfalse
   \leftline{\twelvebold Figure Captions}\par
   \nobreak\vskip\headskip \penalty 30000
   \input figures.aux
   }
%%%%%%%%%%%%%%%%%%%%%%%%%%%%%%%%%%%%%%%%%%%%%%%%%%%%%%%%%%%%%%%%%%%%%%
% Text starts here
%%%%%%%%%%%%%%%%%%%%%%%%%%%%%%%%%%%%%%%%%%%%%%%%%%%%%%%%%%%%%%%%%%%%%%
\pubnum{471}
\secondpubnum{hep-ph/9411297}
\Month{November 1994}
\titlepage
\title{\begingroup\textfont1=\seventeenbi \scriptfont1=\twelvebi
\scriptfont0=\twelvebf \scriptfont2=\twelvebsy
\seventeenbf Four-fermion decay of Higgs bosons\endgroup}
\author{\fourteenrm T.~Asaka and Ken-ichi Hikasa}
\address{Physics Department, Tohoku University\break
Aoba-ku, Sendai 980-77, Japan}

\abstract

We calculate four-fermion decays of a Higgs boson via $WW$ and/or $ZZ$
intermediate states for Higgs masses below $m_W$.
We examine models with a doubly-charged Higgs boson $H^{++}$
and show that the four-fermion decay is the dominant mode for a wide range of
parameter space.  Existing searches for $H^{++}$ in $Z$ decays have not looked
for this mode.
We also derive four-fermion decay rate for a neutral Higgs boson.

\endpage

\chapter{Introduction}

Although the standard model has been very successful in describing the
interaction of elementary particles,
all available tests are sensitive to its gauge part only.
The Higgs sector of the theory remains totally untested.
In the minimal standard model, the Higgs sector consists of a single doublet
of Higgs fields, while two doublets exist
in the minimal supersymmetric standard model.
Although the doublet is the simplest possibility and is necessary
to generate quark and lepton masses,
there is no {\it a priori\/} reason that only Higgs doublets exist.
In many extensions of the standard model, larger Higgs representations
appear\ref{For a review, see J.~F. Gunion, H.~E. Haber, G.~L. Kane,
and S.~Dawson, {\sl The Higgs Hunter's Guide\/} (Addison-Wesley,
Redwood City, USA, 1990).}.
A characteristic particle common to these extensions is
a doubly charged Higgs boson, which is not contained in models with doublets.

The smallest Higgs representation%
\footnote{We define Higgs field as a scalar
field with nonzero vacuum expectation value.  Our normalization of the
weak hypercharge is $Q=I_3+Y$.}
containing a doubly charged Higgs is
the triplet representation of SU(2)  with hypercharge $Y=1$.
Such a triplet can generate a Majorana mass for neutrinos
by a Yukawa coupling.  The doubly charged Higgs boson $H^{++}$ can decay to a
same-sign charged lepton pair by this Yukawa coupling.

An important constraint on an extended Higgs sector comes from
the so-called rho parameter\ref{M.~Veltman, \NPB123, 89 (1977).} defined by
$$ \rho = {m_W^2\over m_Z^2\cos^2\!\theta_W} \;, \Eq $$
which is experimentally equal to one in a percent level.
Models with only doublets satisfy $\rho=1$ at the tree level, but those
with an extended Higgs sector gives $\rho\neq1$ in general.  The $Y=1$
triplet alone gives $\rho=1/2$.
The vacuum expectation value (vev) of the triplet has thus to be much smaller
than the vev of the doublet.

A different possibility to give $\rho=1$ has been proposed by Georgi {\it et
al.}{}\ref{H.~Georgi and M.~Machacek, \NPB262, 463 (1985);\nextline
P.~Galison, \NPB232, 26 (1984);\nextline
R.~S. Chivukula and H.~Georgi, \PL182B, 181 (1986).}
and Chanowitz and Golden%
\ref{M.~S. Chanowitz and M.~Golden, \PL165B, 105 (1985).}.
They introduced another triplet with
$Y=0$ and showed that $\rho=1$ results if the two triplets have equal
vev.  Chanowitz and Golden wrote down a Higgs potential with a tree-level
$\SU(2)_L\times\SU(2)_R$ symmetry to ensure $\rho=1$.  In this model, which
we will call the tri-triplet model,
the vev of the triplet can be as large as the vev of the doublet.

Search for $H^{++}$ in $Z$ decay has been
performed by Mark II Collaboration at SLC%
\Ref\MarkII{M.~Swartz \etal, \PRL64, 2877 (1990).}
and OPAL Collaboration at LEP\Ref\OPAL{OPAL Collaboration, P.~D. Acton \etal,
\PLB295, 347 (1992).}{}.
The $H^{++}$ can be pair produced in $Z$ decay with a sizable branching
ratio if energetically
allowed.  They looked for same-sign lepton pairs coming from $H^{++}$ decay.
If the Yukawa coupling is very small, the lifetime of $H^{++}$ becomes
long enough that the particle does not decay in the detector.
The OPAL Collaboration searched for doubly charged heavy particle
tracks to cover this case.  These searches have excluded a doubly charged
Higgs almost up to the kinematical limit ($m_Z/2$) for most values of the
Yukawa coupling.

In these searches, it is assumed that the same-sign lepton pair is the
sole decay mode of $H^{++}$.  However, it is possible that $H^{++}$
decays to four fermions via a pair of virtual $W^+$.
In this paper, we show that this decay mode can be the dominant mode
of $H^{++}$, especially when the triplet vev is not small.  The existing
searches are thus not complete.

We also calculate the four-fermion decay rate of a neutral Higgs boson
via $WW$ and $ZZ$ intermediate states.

%Its coupling to $Z$ is given by  $g_Z(I_3 - 2\sin^2\theta_W)$,

%%%%%%%%%%%%%%%%%%%%%%%%%%%%%%%%%%%%%%%%%%%%%%%%%%%%%%%%%%%%%%%%%%%%%%%%%%%%
\chapter{$H^{++} \to f_1\bar f_2 f_3 \bar f_4$}

The interaction Lagrangian of $H^{++}$ with two $W$'s is derived from the
Higgs kinetic term with covariant derivative, after replacing the neutral
Higgs field by its vev.  It can be written as
$$ {\cal L} = \half g m_W^@ x \, H^{++} W_\mu W^\mu \;,\Eq$$
where $g$ is the SU(2) gauge coupling, $m_W$ is the $W$ mass, and $x$
is a constant depending on the representation and vev of the Higgs field
in which $H^{++}$ is contained.

For the $Y=1$ triplet
$$\chi = \pmatrix{\chi^{++}\cr\chi^+\cr\chi^0\cr} \Eq$$
one finds
$$ x = {2\sqrt2\,v_3\over v} \;,\Eq$$
with $\VEV{\chi^0} = v_3/\sqrt2$, and the normalization $v\simeq246\GeV$ is
defined by
$m_W=\half gv$.  In models with a $Y=1$ triplet and one or more doublets,
we have $v^2 = v_2^2 + 2v_3^2$,  where $v_2$ is the vev of the
standard Higgs doublet or the sum of vev squared for the doublets.
As discussed earlier, $v_3$ is subject to a constraint
from $\rho$.   A recent analysis\ref{P.~Langacker and M.~Luo, \PRD44, 817
(1991).} gives $v_3<25\GeV$,  which is translated to $x<0.29$.

The tri-triplet model contains an additional $Y=0$ triplet.  Though the
coupling is the same, $v^2$ is now given by $v^2=v_2^2+4v_3^2$.  The coupling
is maximal when the triplet vev is dominant, in which case $x=\sqrt2$.
Although the doublet vev cannot be too small (otherwise the top Yukawa
coupling becomes strong), $x\sim1$ is still allowed.%
\REF\Asaka{T.~Asaka, Master Thesis (in Japanese), Tohoku University (1994).}%
\footnote{The rare decay $b\to s\gamma$ constrains the Yukawa coupling
as a function of $H^+$ mass.  Since cancellation occurs in the
tri-triplet model, $v_2$ as small as 100\GeV\ is consistent with the data
even for a $H^+$ mass around 50\GeV\refmark{\Asaka}.}
The $HWW$ coupling can thus be quite sizable in both models.

Let us consider the process
$$H^{++}(q) \to f_1 (p_1) + \bar f_2 (p_2) + f_3 (p_3) + \bar f_4 (p_4) \Eq$$
via this coupling (the particle momenta are in the parenthesis).
We are interested in the mass range %$M \equiv m(H^{++}) < m_W^@$,
$ \mH < m_W^@$, in which decays to an on-shell $W$,
$H^{++}\to W^+W^+$ or $H^{++}\to W^+ f\bar f'$ are kinematically forbidden.
We assume $f_1\neq f_3$, $\bar f_2\neq \bar f_4$ for a while and
neglect the Kobayashi-Maskawa mixing.
The Feynman graph for this process is shown in Fig.~1.
\FIG\?{Feynman diagram for the process $H^{++}\to f_1\bar f_2 f_3 \bar f_4$.}
The amplitude is
$$\eqalign{{\cal M} &= -{1\over8}\,x g^3 m_W\,
{1\over \bigl[m_W^2-(p_1+p_2)^2\bigl] \bigl[m_W^2-(p_3+p_4)^2\bigl]} \cr\r
&\qquad\times \bar u(p_1)\gamma_\mu(1-\gamma_5)v(p_2)\;
\bar u(p_3)\gamma^\mu(1-\gamma_5)v(p_4) \;.\cr}\Eq$$
The decay rate for massless fermions is calculated to be
$$\Gamma(H^{++}\to f_1\bar f_2 f_3 \bar f_4)
%= {\sqrt2 x^2 \GF^3 \mH^7 \over 1152\pi^5}\;
= {x^2 \GF^3 \mH^7 \over 576\sqrt2\,\pi^5}\;
I\Bigl({\mH^2\over m_W^2}\Bigr) \;,\Eqn\eqGamma$$
with the function $I(r)$ defined as
$$ I(r) = \int\!\!\!\int dx\,dy \,{\bar\beta(x,y)\,\bigl[(1-x-y)^2+8xy\bigr]
\over (1-rx)^2(1-ry)^2} \;,\Eqn\eqIr$$
where the integration is over the region $x$, $y>0$, $\sqrt x+ \sqrt y<1$,
and
$$ \bar\beta(x,y) = \bigl[ 1- 2(x+y) + (x-y)^2\bigr]^{1/2} \;.\Eq$$
In the light Higgs limit $r\to0$, the integral may be done analytically:
$$ I(0) = {1\over20} \;.\Eqn\eqIzero$$

We now turn to the case in which the two virtual $W$'s decay to the same
fermion pair, $f_1=f_3$, $\bar f_2=\bar f_4$.  There are two Feynman
graphs for this process as shown in Fig.~2.
\FIG\?{Feynman diagrams for the process $H^{++}\to f_1\bar f_2 f_1 \bar f_2$.}
The amplitude after Fierz rearrangement is
$$\eqalign{{\cal M} &= -{1\over8}\,x g^3 m_W\,
\biggl\lbrace
{1\over \bigl[m_W^2-(p_1+p_2)^2\bigl] \bigl[m_W^2-(p_3+p_4)^2\bigl]} \cr\r
&\qquad\qquad\qquad\quad
+ {1\over \bigl[m_W^2-(p_1+p_4)^2\bigl] \bigl[m_W^2-(p_3+p_2)^2\bigl]}
\biggr\rbrace \cr\r
&\qquad\times \bar u(p_1)\gamma_\mu(1-\gamma_5)v(p_2)\;
\bar u(p_3)\gamma^\mu(1-\gamma_5)v(p_4) \;.\cr}\Eq$$
In the light Higgs limit, we can easily see that the two contributions
have perfectly constructive interference.  The decay rate thus gains a
factor of 4, but the total rate has to be multiplied by 1/4 to account for
the two identical particle pairs in the final state.  The final result in
this limit thus turns out to be the same\footnote{The case
$f_1=f_3$ but $\bar f_2\ne \bar f_4$ (e.g. $u\bar du\bar s$ final state with
identical colors) appears if the Kobayashi-Maskawa mixing is retained.
The amplitude has two contributions with constructive interference, but the
identical particle factor is only 1/2.  The net result is a factor 2 increase
compared to those discussed in the text.}
as Eq.~\eqGamma.

As the Higgs mass increases, the interference of the two amplitudes becomes
less complete.  The two contributions becomes totally incoherent
above the $WW$ threshold, where the width for the identical fermions
is relatively suppressed by a factor of 2.  This can be easily understood
because two channels exist for the different fermion final state
($W_1\to f_1\bar f_2$, $W_2\to f_3\bar f_4$ and
$W_1\to f_3\bar f_4$, $W_2\to f_1\bar f_2$) whereas only one channel is
possible for the identical final state
($W_1\to f_1\bar f_2$, $W_2\to f_1\bar f_2$).

Taking $e^+\nu_e$, $\mu^+\nu_\mu$, $\tau^+\nu_\tau$, $u\bar d$
and $c\bar s$ pairs as the final states (with due attention to the color
factor), we find the total rate for $H^{++}\to f\bar ff\bar f$
$$\Gamma(H^{++}\to f\bar ff\bar f)
= (36 + 9\zeta) {x^2 \GF^3 \mH^7 \over 576\sqrt2\pi^5}\;
I\Bigl({\mH^2\over m_W^2}\Bigr) \;,\Eqn\eqGammaffff$$
where $1/2\leq\zeta\leq1$ and $\zeta\to1$ at $\mH\to 0$.  Since we are not
concerned with an error of order 10\%, we will use $\zeta=1$ in
the following analysis.

The decay width \eqGammaffff\ for $x=1$ is plotted in Fig.~3 as a function
of the Higgs mass.  The pointlike approximation using \eqIzero\ works
quite well for $\mH \lsim 50\GeV$: The difference from the exact result
is less than 20\%.

\FIG\?{Decay width for $H^{++}\to f\bar f f\bar f$.
Three generations of quarks and leptons except $t\bar b$ are included in
the final states.  The solid line shows the numerical result with \eqIr,
whereas the dotted line is the pointlike approximation with \eqIzero.}

%%%%%%%%%%%%%%%%%%%%%%%%%%%%%%%%%%%%%%%%%%%%%%%%%%%%%%%%%%%%%%%%%%%%%%%%%%
\chapter{$H^{++}$ Lifetime and Branching Ratio}

\REF\Hplus{%
ALEPH Collaboration, D.~Decamp \etal, \PRep216, 253 (1992);\nextline
L3 Collaboration, O.~Adriani \etal, \PLB294, 457 (1992).}

Another important decay of $H^{++}$ is the same-sign dilepton mode,%
\footnote{We do not consider decay modes with a Higgs in the
final state, e.g., $H^{++}\to H^+ f\bar f'$ and $H^{++}\to H^+H^+$,
because the LEP limit\refmark{\Hplus} $m(H^+)\gsim42\GeV$ makes these decays
unlikely to be important.}
which has been discussed in the literature%
\ref{See e.g., J.~A. Grifols, A.~M\'endez, and G.~A. Schuler, \MPLA4, 1485
(1989);\nextline
M.~L. Swartz, \PRD40, 1521 (1989).}.
Here we rederive the results to fix our notation.

The $Y=1$ triplet can couple to a pair of left-handed leptons.%
\footnote{Besides the $Y=1$ triplet, $Y=2$ singlet (which appear in
left-right symmetric models) can couple to a (right-handed) charged lepton
pair.  These two exhausts possible Higgs representations which can decay to
$\ell^+\ell^+$.}
We write the Yukawa coupling as
$$\eqalign{{\cal L} &= -\half \tsum_{i,\,j} h_{ij} \bar L^c_i \tau^a \chi^a L_j
+ \hbox{h.c.} \cr
&= -{1\over\sqrt2} \tsum_{i,\,j} h_{ij}
\bigl( \bar\ell_i^c \ell_{jL}\,\chi^{++}
+ \sqrt2\,  \bar\nu_i^c \ell_{jL}\, \chi^+
+ \bar\nu_i^c \nu_{jL}\, \chi^0 \bigr) + \hbox{h.c.}  \;,\cr}\Eq$$
where $i$, $j$ are the generation indices, $h_{ij}$ forms a symmetric Yukawa
coupling matrix, and
$$\eqalignno{ & L_i = \pmatrix{\nu_i\cr \ell_i\cr}_L \;,\qquad
L_i^c = i\tau_2 C\bar L_i^T = \pmatrix{\ell^c_i\cr -\nu^c_i\cr} \;,&\Eq\cr\r
& \tau^a\chi^a = \pmatrix{ \chi^+ & \sqrt2\chi^{++}\cr -\sqrt2\chi^0
& -\chi^+\cr} \;. &\Eq\cr}$$
The neutrino mass matrix is given by $(m_\nu)_{ij} = h_{ij} v_3 $.
The dilepton decay rate is for an identical lepton pair
$$\Gamma(H^{++} \to \ell^+\ell^+) = {|h_{\ell\ell}|^2\over 16\pi}\,\mH \;,\Eq$$
and for a different lepton pair $\ell\neq\ell'$
$$\Gamma(H^{++} \to \ell^+\ell'{}^+) = {|h_{\ell\ell'}|^2\over 8\pi}\,\mH
\;.\Eq$$

The total decay rate is given by the sum of Eq.~\eqGammaffff\ and these
dilepton decay rates.  For definiteness, we represent the dilepton decays
by that with largest $h_{\ell\ell'}$, which we take to be $\tau\tau$.
The lifetime is thus
$$ \tau^{-1} = \Gamma(f\bar ff\bar f) + \Gamma(\tau\tau) \;.\Eq$$

The $H^{++}$ can be pair produced in $Z$ decay if kinematically allowed.
The partial width is
$$\Gamma(Z\to H^{++} H^{--}) = {\GF m_Z^3\over 6\sqrt2\pi}\,
(I_3-2\sin^2\!\theta_W)^2 \beta^3 \;,\Eq$$
where $I_3=2-Y$, $\beta=(1-4\mH^2/m_Z^2)^{1/2}$.

The experimental signature of $H^{++}$ production
depends on the lifetime and dominant decay mode.  There are three distinct
regions depending on $\mH$, $x$, and $h_{\tau\tau}$, as is demonstrated
in Fig.~4 for $\mH=40\GeV$.

\REF\ARGUS{ARGUS Collaboration, H.~Albrecht \etal, \PLB292, 221 (1992).}
\REF\Kolb{E.~W. Kolb and M.~S. Turner, {\sl The Early Universe\/}
(Addison-Wesley, Redwood City, USA, 1990).}

\FIG\?{Decay signatures of $H^{++}$ for $\mH=40\GeV$.
The solid lines correspond to the ratio of four-fermion to dilepton
branching fractions $R=0.1$, 1, and 10.  The dash-dotted lines give the
contour for the decay length $\ell=1$ cm and 1 m.
The laboratory limit\refmark{\ARGUS}, $m(\nu_\tau)<31\MeV$,
and cosmological limit\refmark{\Kolb} for a stable $\nu$, $m(\nu)<50\eV$,
are also shown by the dashed lines.}

First, if the decay length
$\ell=\gamma\beta c\tau$ is larger than the size of the charged track
detector, an event looks
like a pair of doubly charged stable particles.  This happens in the lower
left side of Fig.~4, in which
lines corresponding to $\ell=1\m$ and $\ell=1\cm$ are shown.

If $\ell$ is sufficiently small that the decay products are visible in
the detector, the signature depends on the dominant decay mode.  We define
the ratio
$$ R = {\Gamma(H^{++}\to f\bar f f\bar f) \over \Gamma(H^{++}\to\tau^+\tau^+)}
\;\Eq$$
which measures the relative importance of the two decay modes.
Lines corresponding to $R=0.1$, 1, and 10 are shown in Fig.~4.
In the region both $R\ll1$ and $\ell$ are small,
an event contains four leptons.  This is the case if the Yukawa coupling
is rather large and the triplet vev is small.

Finally, if $\ell$ is small and $R\gg1$, the four-fermion mode dominates and
an event contains eight quarks/leptons in the final state.  This can
occur in the lower right part of the plot.  This region grows rapidly
for a larger Higgs mass because of the $\mH^7$ dependence of the width.

As mentioned earlier, the first two signatures have been sought
for\refmark{\MarkII, \OPAL}, but the last possibility has hitherto
overlooked.  In this case, a mass limit for $H^{++}$ can still be derived from
the total $Z$ width.  Reference [\OPAL] gives a limit of 30.4\GeV\ (95\%CL)
for $Y=1$.  Direct search for the eight-fermion final states would
improve the bound up to the kinematical limit of 45\GeV.

%%%%%%%%%%%%%%%%%%%%%%%%%%%%%%%%%%%%%%%%%%%%%%%%%%%%%%%%%%%%%%%%
\chapter{Four-fermion Decay of Neutral Higgs Boson}

In general, a neutral Higgs boson $H^0$ couples with a $W$ and $Z$ pair:
$$ {\cal L} = g m_W^@ x_c W^\dagger_\mu W^\mu H^0
+ \half g_Z^@ m_Z^@ x_n Z_\mu Z^\mu H^0 \;,\Eq$$
where $g_Z^@=g/\cos\theta_W$.  In the standard model, we have
$x_c=x_n=1$ for the neutral Higgs boson.
These couplings induce the four-fermion decay mode
$H^0\to f_1 \bar f_2 f_3 \bar f_4$.  The decay width can
be obtained in a similar way as for $H^{++}$.
There are two possible intermediate states $W^{+*}W^{-*}$ and $Z^*Z^*$.
Some of the final states (e.g.\
$e^+e^-\nu_e\bar\nu_e$) receive interefering contribution from both
intermediate states.  Another set of final states like $e^+e^-e^+e^-$
have two diagrams with $Z^*Z^*$ intermediate state.  Taking these
complications in account, we find the decay rate for $m_H\ll m_W$
$$\eqalign{\Gamma(H^0\to f\bar f f\bar f)
&= {\GF^3 m_H^7 \over 11520\sqrt2\,\pi^5}\;\cr
&\quad\times\biggl\lbrace
81 x_c^2
+ \bigl[ 12(v_\nu{+}a_\nu)(v_e{+}a_e) + 24(v_u{+}a_u)(v_d{+}a_d) \bigr]
x_c x_n \cr
&\qquad + \biggl[ 12\tsum_{f}k_f\bigl( v_f^4{+}6v_f^2a_f^2{+}a_f^4
\bigr) %\cr&\qquad\qquad\qquad
+ 72 \Bigl( \tsum_{f}k_f\bigl( v_f^2{+}a_f^2 \bigr)
\Bigr)^2\biggr] x_n^2 \;\biggr\rbrace \;,\cr}\Eq$$
where the sum fur $f$ runs over $f=\nu$, $e$, $u$, $d$,
$$ k_f = \cases{1 & $f=\nu$, $e$, \cr 2 & $f=u$,\cr 3 & $f=d$,\cr}\Eq$$
and $v_f=\tfrac12 I_{3}(f_L) - Q_f \sin^2\!\theta_W$,
$a_f=\tfrac12 I_{3}(f_L)$.
The expression within $\{\}$ is equal to 103.8 for $\sin^2\!\theta_W=0.23$,
$x_c=x_n=1$, most of which comes from the $WW$ intermediate state.
The decay rate is much smaller than that of $H^0\to f\bar f$ via the Yukawa
coupling.  This result may, however, be relevant for a neutral Higgs in a
nondoublet representation, which cannot couple to a fermion pair.  Such
a particle has been searched for by ALEPH Collaboration at LEP{}%
\Ref\ALEPH{ALEPH Collaboration, D.~Decamp \etal, \PLB262, 139 (1991).}.
Our result is in disagreement with the calculation quoted there.%
%\ref{R.~Kleiss, quoted in Ref.~[\ALEPH].}.

%%%%%%%%%%%%%%%%%%%%%%%%%%%%%%%%%%%%%%%%%%%%%%%%%%%%%%%%%%%%%%%%%%%%%%%
\chapter{Summary}

Various models contain a doubly charged Higgs boson.  A particularly
interesting class of models have a $Y=1$ triplet Higgs, which can give
a Majorana mass to neutrinos.  The left-right symmetric model is an
example of such a model.  Although the triplet vev is constrained by the
$\rho$ parameter, its magnitude may be still substantial.  In this case,
the doubly charged Higgs decays dominantly to four-fermion final states
via intermediate $WW$.  Existing searches for $H^{++}$ in $Z$
decays only looked for either dilepton modes or quasistable massive tracks.
A new analysis is thus called for to exclude the entire mass region
$m_H^@<m_Z^@/2$.

We have also derived the four-fermion decay rate for a neutral Higgs boson.
This decay mode may be important for a special class of Higgs which has
a reduced Yukawa coupling to quarks and leptons.

\endpage
%%%%%%%%%%%%%%%%%%%%%%%%%%%%%%%%%%%%%%%%%%%%%%%%%%%%%%%%%%%%%%%%%%%%%%%%
\par \penalty-400 \vskip\chapterskip
   \spacecheck\referenceminspace \immediate\closeout\referencewrite
   \referenceopenfalse
   \leftline{\twelvebold References}\par
   \nobreak\vskip\headskip \penalty 30000
   \input reference.aux
   \endpage
\nopagenumbers
\par \penalty-400 \vskip\chapterskip
   \spacecheck\referenceminspace \immediate\closeout\figurewrite
   \figureopenfalse
   \leftline{\twelvebold Figure Captions}\par
   \nobreak\vskip\headskip \penalty 30000
   \input figures.aux

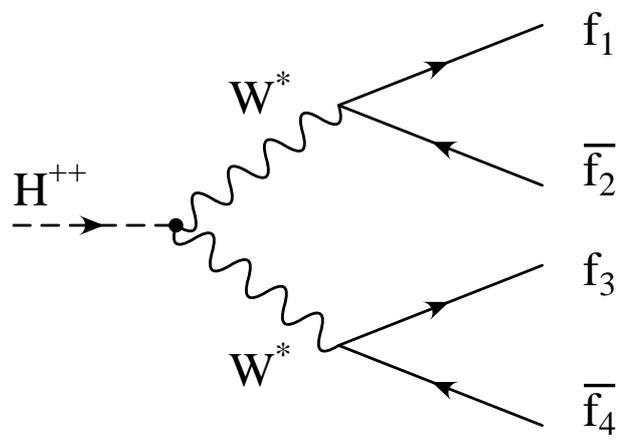
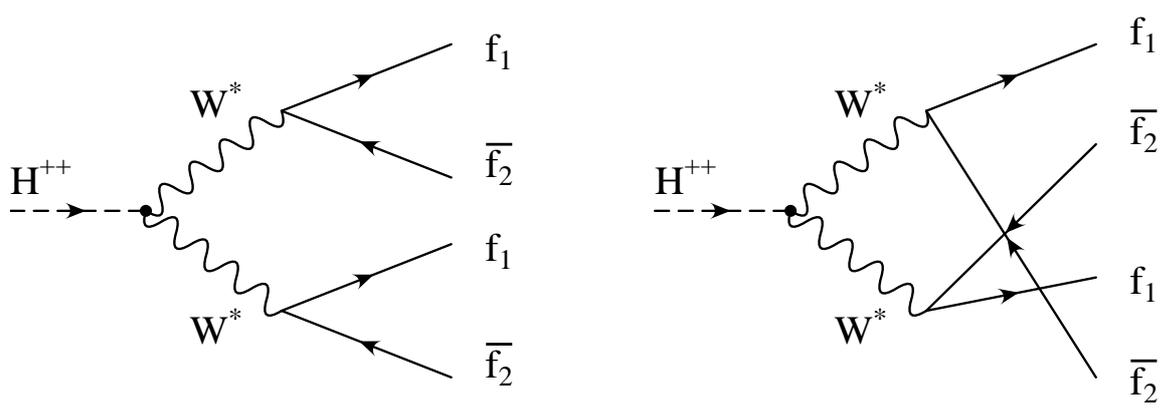
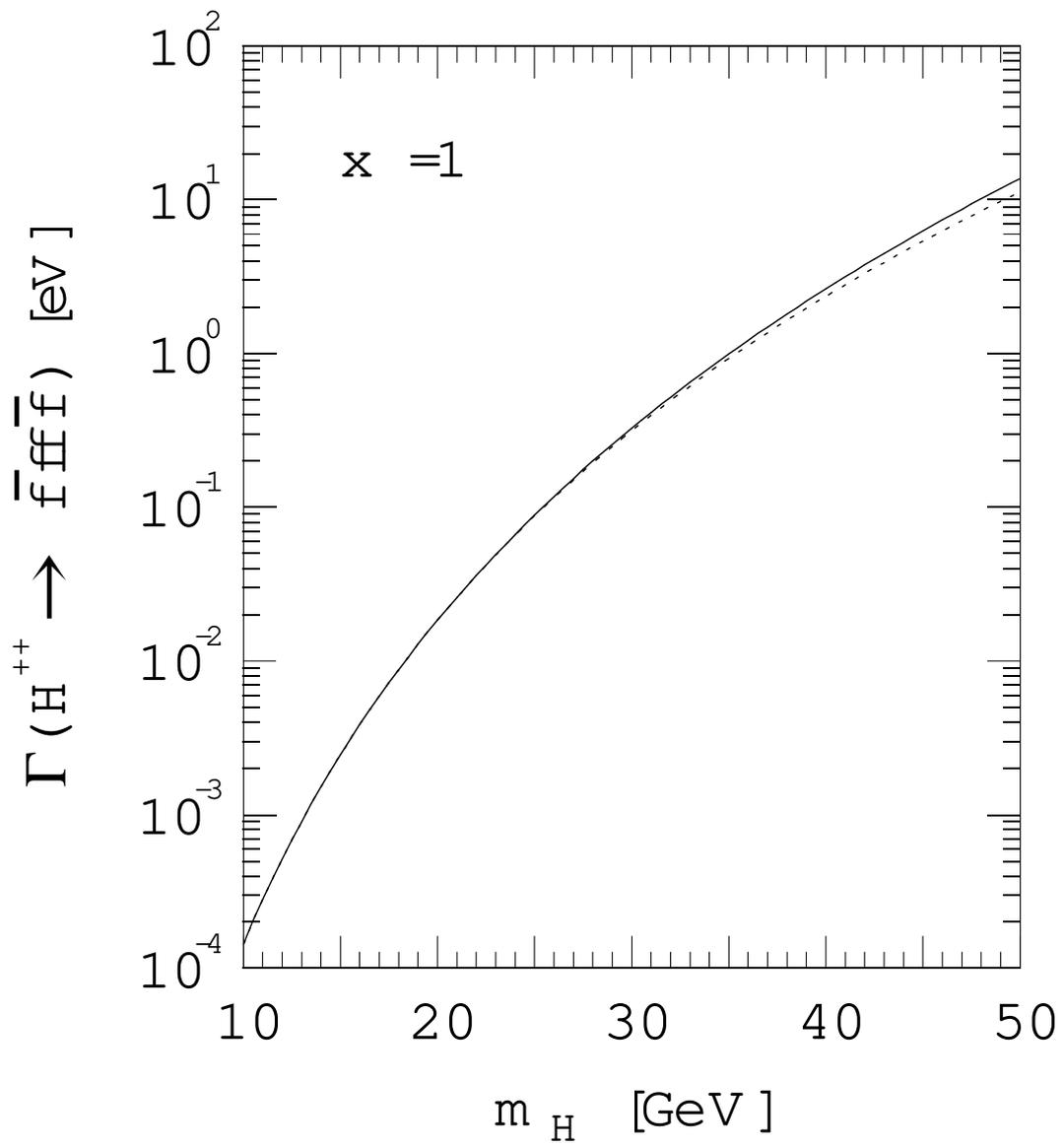
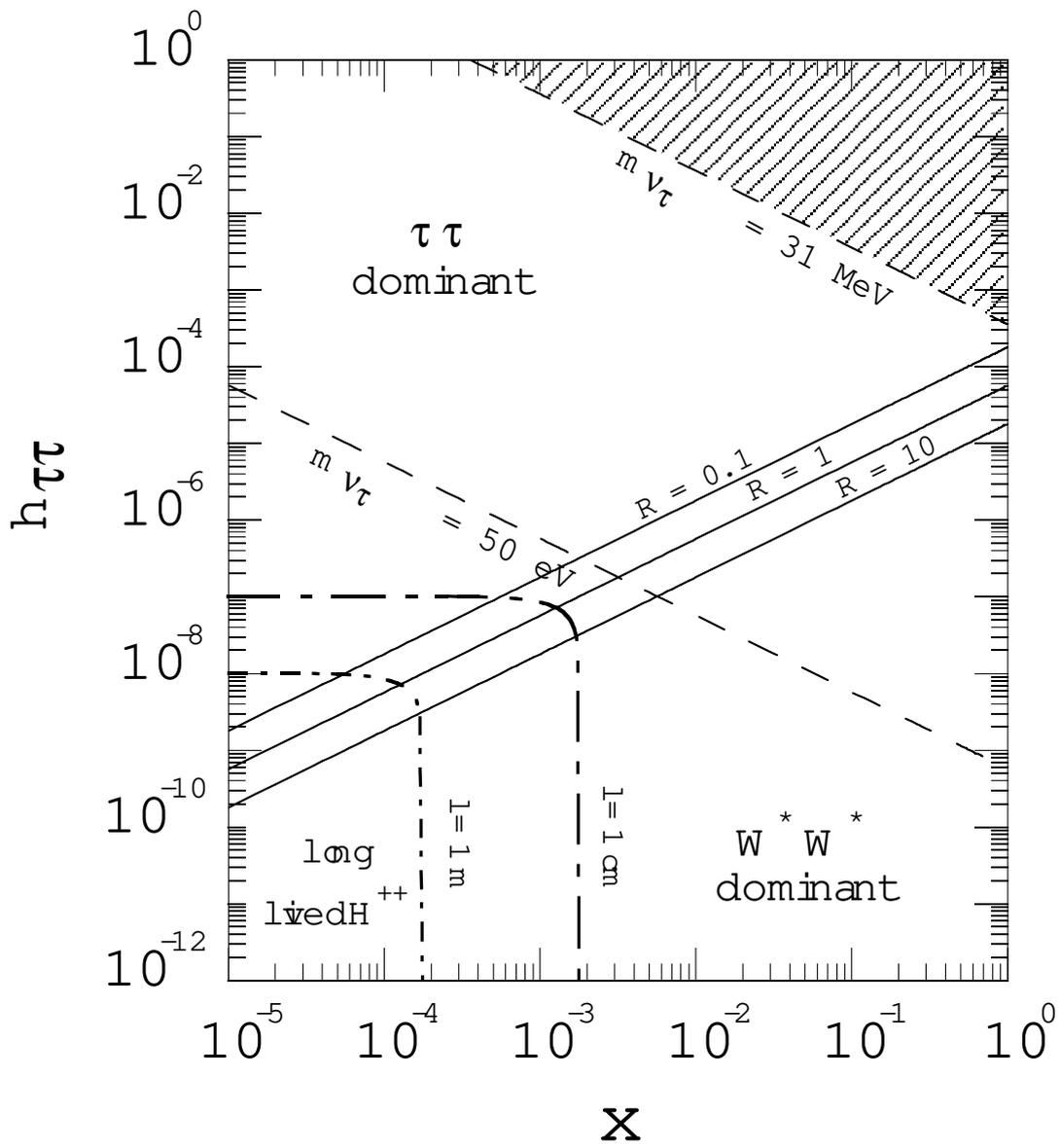
\bye